\documentclass[pra,aps,amsmath,twocolumn,floatfix]{revtex4-1}
\usepackage{amsmath,graphicx}
\begin{document}
\def\tr{\rm{Tr}}
\def\la{{\langle}}
\def\ra{{\rangle}}
\def\a{{\alpha}}
\def\e{\epsilon}
\def\q{\quad}
\def\w{\tilde{W}}
\def\t{\tilde{t}}
\def\a{\hat{A}}
\def\h{\hat{H}}
\def\E{\mathcal{E}}
\def\p{\hat{P}}
\def\u{\hat{U}}
\def\n{\\ \nonumber}
\def\j{\hat{j}}
\def\alph{a}
\def\vc{\underline{c}}
\def\vf{\underline{f}}
\title{Reply to L.Vaidman's comment on "Asking photons where they have been in plain language"  }
\author {D. Sokolovski$^{a,b}$}
\affiliation{$^a$ Departmento de Qu\'imica-F\'isica, Universidad del Pa\' is Vasco, UPV/EHU, Leioa, Spain}
\affiliation{$^b$ IKERBASQUE, Basque Foundation for Science, E-48011 Bilbao, Spain}
\date{\today}
\begin{abstract}
Vaidman's analysis of the photon's past is based on incorrect interpretation 
of quantum probability amplitudes. The confusion stems from his original 
work  [{\it Phys. Rev. Lett.}   \textbf{60}, 1351 (1988)], which missed the connection between the amplitudes and the  "weak values" introduced therein. 
\end{abstract}
\maketitle
\noindent
{Keywords; {\it quantum particle's  past, transition amplitudes, weak measurements}}
\vspace{0.5cm}

Recently Lev Vaidman has presented his Comment \cite{COMM} on our paper \cite{PLA2017}, in which we 
argued that the description the photon's past in \cite{F1}, based on "weak trace" approach of \cite{v2013},
is incorrect. Vaidman disagreed \cite{COMM}, and this is our reply to the Comment.
\newline
First we note that the problem, formulated in \cite{v2013} and elaborated in \cite{F1},\cite{F2}, is an artificial one.
A wave packet injected into the interferometer shown in Fig.1 will split at the first beam splitter (BS1) into two parts, one of which will go into BS4, and split again so that one part will go to the detector $D$, and the other will leave the system. The part which took the left turn at BS1  will reach BS2, and split again. The two new parts will recombine at BS3, and leave the interferometer, prepared in such a way that nothing will pass through the arm connecting BS3 with BS4. The only part of the wave function reaching the detector will have passed through the point C. The reader may read no further, wondering, perhaps, what the argument is about.
\begin{figure}
	\centering
		\includegraphics[width=8.5cm,height=8cm]{{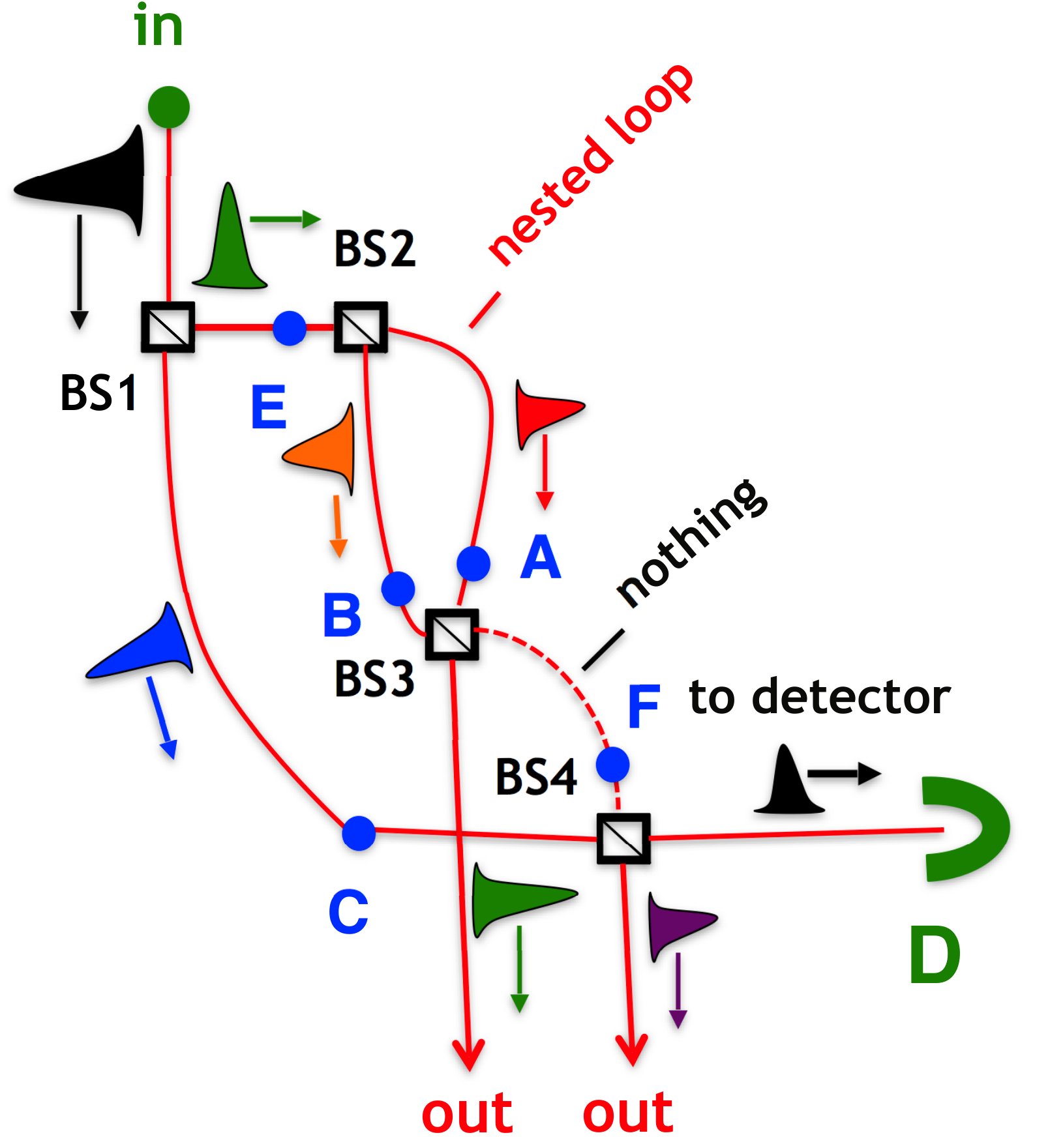}}
\caption{(Color online) 
The relevant part of the interferometer used in \cite{F1}. Also shown is the evolution of a wave packet, 
passing through four beam splitters ($BS$). The parameters are such that no wave function passes through the arm 
$BS3-BS4$.} 
\label{fig:4b}
\end{figure}
\newline
It is about the oversight in the paper \cite{Ah1}, which started the fashionable today field of "weak measurements", and which led Vaidman to conclude that a photon arriving at the detector should have been present in the loop between BS2 and BS3, 
which it could neither enter, nor leave \cite{v2013}. 
Conventionally \cite{FeynL}, a quantum system making a transition between the initial and final states, $|\psi\ra$ and $|\phi\ra$,
is described by a probability amplitude,  $A$, such that $P=|A|^2$ gives the probability of the transition. If there are several ways
(paths) to reach $|\phi\ra$, e.g., different arms of the interferometer leading to the detector, there are several amplitudes, $A_i$, 
and $P=|\sum_i A_i|^2$. The authors of \cite{Ah1} have shown that the mean reading of a meter set to perturb the measured system as little as possible is expressed in terms of the normalised (relative) probability amplitudes $\alpha_j=A_j/\sum_i A_i$. For example, if the measured quantity takes the values $B_i$, the mean reading (shift of the pointer) is $ \sum_iB_i\alpha_i$. If, in order to obtain a yes/no answer to the question "did the system take the $j$-th path?", one chooses a projector $B_i=B_j\delta_{ij}$, he/she will be measuring the amplitude $\alpha_j$. The oversight in \cite{Ah1} consists of not recognising the amplitudes as such, and naming them, or their combinations, "weak values" instead. Along with the new name, the "weak values" acquired also new predictive powers, not enjoyed by the probability amplitudes in quantum mechanics. 
\newline
The weak measurements approach of \cite{Ah1} is responsible for such exotic predictions as "negative kinetic energy" \cite{NKE}, "negative number of particles" \cite{AhHARDY}, "having one particle in several places simultaneously" \cite{AhBOOK}, "photons disembodied from their polarisations" \cite{CAT}, 
"electrons with disembodied charge and mass" \cite{CAT}, and "atoms with the internal energy disembodied from the mass" \cite{CAT}.
While an over-sympathetic reader might see them as mere exaggerations \cite{PLA2016},  the idea that photons can be "found in places they neither enter or leave" \cite{v2013}, \cite{F1} is  just incorrect, as was pointed out by several authors in Refs. \cite{WI}-\cite{WF}.  

 {In his Comment \cite{COMM} Vaidman suggests that our critique \cite{PLA2017} relies on the concept of Feynman paths
whose "ontological meaning" is shared by "only a small minority" and which are "everywhere". 
In fact, we rely only on the general principle formulated in a standard (more than 20 editions since 1964) undergraduate textbook 
\cite{FeynL} as: "When an event can occur in several alternative ways, the probability amplitude for the event  is the sum of the probability amplitudes for each way considered separately".}
\newline
In Fig.1there are three ways leading to the detector with non-zero amplitudes $\alpha_C$, $\alpha_{EAF}$, and $\alpha_{EBF}$, passing via $C$, $E\to A \to F$, and $E\to  B \to  F$, respectively. 
\begin{figure}
	\centering
		\includegraphics[width=6cm,height=6cm]{{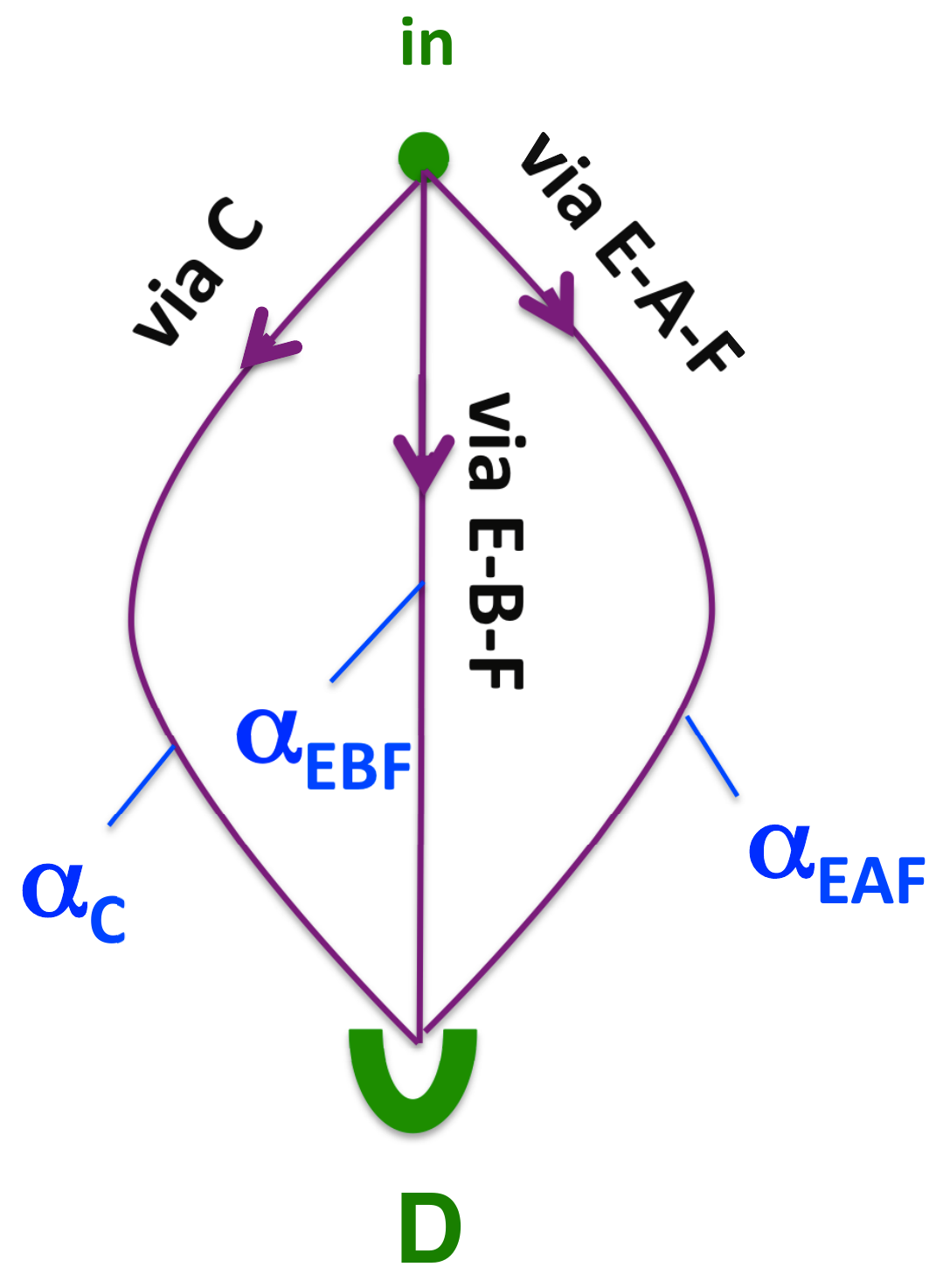}}
\caption{(Color online) 
Virtual paths, and the corresponding (relative) probability amplitudes, available for a photon which ends up in the detector $D$.
The parameters are such that $\alpha_{EBF}=-\alpha_{EAF}$.} 
\label{fig:4b}
\end{figure}
In his "weak trace" approach, Vaidman \cite{v2013} postulates that the photon should be present whenever a weak value of the corresponding projector, measured by a local pointer in nonzero.
Thus, the photon "has  been" at locations $A$, $B$, and $C$, and "was not" at $E$ and $F$.
This is precisely a kind of statement one is not allowed to make in standard quantum mechanics \cite{FeynL}, and the invention of "weak values" in \cite{Ah1}, has done little to change that.
\newline
In reality,  the author of \cite{v2013} has  established something else. None-zero weak traces at $A$, $B$, and $C$ signal only that 
$\alpha_C$, $\alpha_{EAF}$ and $\alpha_{EBF}$ do not vanish. In addition, zero weak traces at $E$ and $F$ indicate that
$\alpha_{EAF}=-\alpha_{EBF}$. This, however, was known from the start, since we built the interferometer in this way. 
\newline
It is also known that this information cannot be translated into the statements about where the photon has and has not been. 
Discussing a closely related Young's double-slit experiment, Feynman gave the following warning to his students \cite{FeynL}.
"But, when one does not try to tell which way the electron
goes, where there is nothing in the experiment to disturb the electrons,
then one may not say that an electron goes either through hole 1 or hole 2. 
If one does say that, and starts making deductions from the statement, he will 
make errors in the analysis."
{Vaidman \cite{COMM} is correct in noticing that the formalism of standard quantum mechanics "has no concept for the past of pre- and post-selected particle". The problem is that, according to Feynman, such a concept cannot be had in any meaningful way, 
since errors will always be made.} 
\newline
The error in the analysis of \cite{v2013} became an error in the explanation given in \cite{F1},
where the authors claim to find that "Some of them [photons] have been inside the nested
interferometer ..., but they never entered and never left". 
Since the publication of \cite{F1} several authors  \cite{WI}-\cite{WF}, including ourselves \cite{PLA2017}, have been pointing out 
that in an experimental realisation there is always evidence of photons entering the nested loop between 
$BS2$ and $BS3$ in Fig.1. However, the authors of \cite{F1}, probably influenced by the "weak measurement" ideology 
of \cite{v2013}, chose not to look for the much smaller signals from the probes at $E$ and $F$.
In his Comment \cite{COMM}, Vaidman says that "A tiny
leakage of light in the inner interferometer, which leads
to these (below the noise threshold) signals, is explicitly
mentioned in \cite{F1} and calculated, (Eq.8), " in \cite{v2013}." This is not a particularly convincing excuse, since 
the "tiny leakage" is  the very reason for the presence of the photons in the inner interferometer.
The existence of these below the noise signals is vital
for the existence of the strong signals from the probes at $A$ and $B$.
To test their conclusions,  the authors of \cite{F1}, may try blocking the $BS1-BS2$ arm of the interferometer. If no photons ever go there, the blockage would not alter anything else in the experiment. It will, however, stop the signals from $A$ and $B$, since then, in Fig.2, we would also have $\alpha_{EAF}=\alpha_{EBF}=0$.
In a similar way, blocking the $BS3-BS4$ arm would eliminate the signals from $A$ and $B$, without altering the rate at which the photons arrive at the detector. 

In summary, the "weak trace criterion for the past of a quantum particle" proposed in \cite{v2013} amounts to using quantum mechanical probability amplitudes, in a manner they they are not supposed to be used \cite{FeynL}. Applied to to the experiment 
described in \cite{F1}, the criterion gives an fallacious explanation for the observed behaviour of the photons.

\end{document}